# Sub-millimetre wave range-Doppler radar as a diagnostic tool for gas-solids systems – solids concentration measurements

Marlene Bonmann[a], Diana Carolina Guío-Pérez[b], Tomas Bryllert[a], David Pallarès[b], Martin Seemann[b], Filip Johnsson[b], Jan Stake[a]

[a] Department of Microtechnology and Nanoscience, Terahertz and Millimetre Wave Laboratory
Chalmers University of Technology, SE-412 96 Gothenburg, Sweden
[b] Department of Space, Earth and Environment, Division of Energy Technology
Chalmers University of Technology, SE-412 96 Gothenburg, Sweden

## Abstract

Current non-intrusive measurement techniques for characterising the solids flow in gas-solids suspensions are limited by the low temporal or low spatial resolution of the sample volume, or in the case of optical methods, by a short range of sight. In this work, a sub-millimetre wave range-Doppler radar is developed and validated for non-intrusive sensing of solids concentrations in a gas-solids particle system with known characteristics. The radar system combines favourable features, such as the ability to see through at optical frequencies opaque materials, to measure the local solids velocity and the reflected radar power with a spatial resolution of a few cubic centimetres over distances of a few metres. In addition, the radar hardware offers flexibility in terms of installation. After signal processing, the output of the radar is range-velocity images of the solids flowing along the radar's line-of-sight. The image frame rate can be close to real-time, allowing the solids flow dynamics to be observed.

While the well-established Doppler principle is used to measure the solids velocity, this paper introduces a method to relate the received radar signal power to the solids volumetric concentrations ($c_v$) of different particulate materials. The experimental set-up provides a steady stream of free-falling solids that consist of glass spheres, bronze spheres or natural sand grains with known particle size distributions and with particle diameters in the range of 50–300 μm. Thus, the values of $c_v$ found using the radar measurements are validated using the values of $c_v$ retrieved from closure of the mass balance derived from the measured mass flow rate of the solids stream and the solids velocity. The results show that the radar system provides reliable measurements of $c_v$, with a mean relative error of approximately 25% for all the tested materials, particle sizes and mass flow rates, yielding values of $c_v$ ranging from $0.2\times10^{-4}$ m$^3$/m$^3$ up to $40\times10^{-4}$ m$^3$/m$^3$ and solids velocities within the range of 0–4.5 m/s. This demonstrates the ability of the radar technology to diagnose the solids flow in gas-solids suspensions using a unique combination of penetration length, accuracy, and spatial and velocity resolution. In future work, the radar technique will be applied to study non-controlled solids flow at a larger scale, and to understand flow conditions relevant to industrial reactor applications, e.g., fluidised bed, entrained flow, and cyclone units.

## *1.* Introduction

In a wide range of industrial processes, the transport and mixing of gas-solids phases play essential roles. These processes range from the pneumatic transport of particulate materials in, for example, food processing [1] and pharmaceutics [2], to the fluidised bed technology used, for example, in combustion plants [3]. Measuring the velocities and concentrations of solids is essential for understanding the mechanisms that govern the solids flow and for validation of numerical modelling approaches [4, 5] that attempt to describe the complex gas-solids dynamics.

Ideally, a measurement technique to diagnose the solids flow in a gas-solids suspension should:
(i) Be non-intrusive, avoiding the insertion of objects that affect the solids flow to be measured.
(ii) Resolve the dynamics of the flow. It should provide sufficiently high resolutions in the time (time-scales longer than 10$^{-1}$ s cover most of the power spectra of fast gas-solids flows (see for example [6]) and space (with particle scale length being the ultimate target) dimensions.
(iii) Provide penetration into the gas-solids suspension so that the flow can also be diagnosed in locations that lack optical access.
(iv) Be sufficiently robust to withstand the harsh erosional (and sometimes corrosive) flow conditions.
(v) Be associated with a low cost of implementation.

Among the non-intrusive techniques currently available, the most commonly used are time-averaged pressure measurements, from which the axial profiles of the cross-sectional average concentrations of solids can be obtained with a spatial resolution that ranges from decimetres to several metres (e.g., [7]). Specific characteristics, such as bubble dynamics and regime changes, can be extracted from time-resolved pressure measurements (e.g., [8], [9], and [10]). Some non-intrusive tomographic techniques applied in gas-solid systems involve x-rays, γ-rays, electrical capacitance, and phase-Doppler anemometers. They can be used for mapping solids flow (velocity, concentration, or both) in narrow bench-scale units with a spatial resolution in the order of a few centimetres [11, 12, 13 and references therein, 14]. Such dynamic tomographic measurements have limited applicability to large-scale units, owing to the higher integrated absorption of the signal and the rapidly increasing number of sensor pairs needed to retain the same level of spatial resolution. In addition, the intended positioning of the sensor pairs is often restricted by geometric constraints, which may reduce the flexibility and accuracy of the technique. Moreover, the measurement accuracy largely depends on the reconstruction algorithm. Direct tracking techniques, such as electrostatic induction sensors or particle image velocimetry (PIV) [15, 12], offer higher spatial resolution (1 mm). However, these techniques are restricted to measurements of flows at moderate temperatures and are limited by the need for an optically free line-of-sight to the measurement volume. Therefore, solids velocity measurements made with electrostatic induction sensors or particle image velocimetry PIV are rarely used for dense flows.

Radar technology combines several of the desired properties listed above, in that it is non-invasive, has long penetration lengths, and excellent velocity and spatial resolutions. Some authors [16, 17] have demonstrated non-intrusive measurements of multi-disperse solids streams using a multi-static dual-frequency (91.5 GHz, 150.3 GHz) radar system. They observed proportionality between the solids mass load and reflected signal power. The radar technique has also been used to measure particle size distributions (PSD). Even though the radar mode cannot measure the velocity of the solids, Baer and co-workers [18] have demonstrated that two 80-GHz frequency-modulated continuous-wave (FMCW) radars can be used to find the cross-sectional average concentration of the solids volume fraction and velocity across a 200-mm-wide conveying installation. The $c_v$ is derived from the time of flight of the radar signal and the velocity from two-point measurements using the correlation between the two radar signals along the solids flow. However, the configurations involving two transmitter-receiver sets and several radars, respectively, require accurate radar alignment, making them less-suitable for performing measurements in industrial units. Furthermore, the spatial resolution is restricted, being defined by the overlap of the two radar beams [16, 17] and across the whole 200-mm tube [18].

Radar systems that utilise only one antenna for transmitting and receiving the radar signals operating in pulsed-FMCW mode offer an alternative to overcome the complication of radar antenna alignment and to allow (besides local solids concentration measurements) for local solids velocity measurements using the Doppler principle. Cooper and Chattopadhyay [19] have described a 680-GHz radar that can simultaneously monitor the distance and velocity of small solids particles in a sandstorm. However, the radar was not adapted for industrial applications.

The recently developed radar technology with sub-millimetre wave frequencies (325–350GHz) [20] offers high spatial resolution and has a relatively compact footprint (40×30×20 cm$^3$). Thus, it allows the performance of measurements while pointing the radar beam in any desired direction and simplifies the installation for industrial measurements. In addition, it operates at higher frequencies than previous radar systems [16, 17, 18]. In the case of the 680-GHz radar used previously [19], this increases the measurement sensitivity for smaller solids particles and increases the spatial resolution. At the same time, the radar operates below optical frequencies (400–750 THz), which means that the radar beam has a greater depth of penetration than optical measurement methods. Therefore, this sub-millimetre wave radar technology is highly promising for non-intrusive monitoring of solids concentrations and velocities, allowing the characterisation of volumes in the order of 10$^{-3}$ m$^3$ (resulting from a beam cross-section in the order of 0.01–0.1 m$^2$ and a spatial resolution along the direction of the beam of 10$^{-2}$ m), with a velocity resolution in the order of 10$^{-2}$ m/s and a time resolution of 10$^{-2}$ s (frame rate in the order of 10–100 Hz). The abilities of FMCW-pulse Doppler radar systems to measure accurately the velocities of objects are generally accepted and, specifically for the radar system used in this work, have been previously reported in the literature [20]. Thus, there is a need to verify the ability of such radar systems to measure accurately the solids concentrations, thereby providing an all-in-one measurement of the solids flux.

The aim of this work was to evaluate the use of radar technology as a diagnostic tool for the characterisation of solids flows. Here, the 340–GHz sub-millimetre wave FMCW-pulse Doppler radar system described previously [20] is used to measure the solids velocity and concentration along a free-falling solids stream with known mass flow rate, stream diameter and solids properties. Radar-based measurements of the solids concentration are compared to their corresponding reference values calculated from closure of the mass balance using the values of the measured mass flow rate and the solids velocity (obtained through the Doppler method and, thus, considered to be reliable). The radar measurement is tested with solids of different sizes, shapes, and material/dielectric constants and with varying solids concentrations.

## 2. Theory

Figure 1 illustrates the general principle of a radar beam intersecting a gas-solids suspension, where the geometrical dimensions of the gas-solids suspension with solids volume concentration $c_v$ exceeds the radar beam-width $d_{radar} \approx R\theta$, where $R$ is the distance (range) to the radar antenna and $\theta$ is the angular beam-width. The radar beam-width $d_{radar}$ defines the cross-range resolution of the radar, whereas the spatial resolution in the propagation direction of the radar beam is determined by the range resolution $\Delta R$, thereby yielding the sample volume $V_{radar} = \pi \Delta R (d_{radar}/2)^2$ for a given range increment. For each $V_{radar}$, the solids velocity distribution and $c_v$ are measured. The solids velocity is indicated by $v$, the Doppler velocity, whereas $v_d$ is measured by the radar and is the projection towards the radar (i.e., radial velocity).

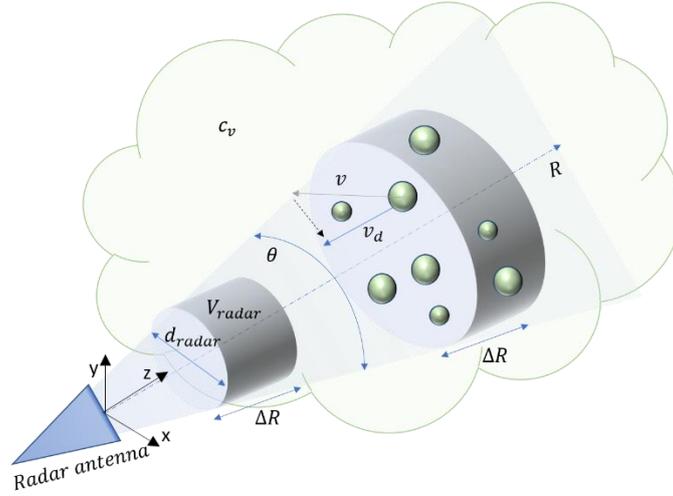

Figure 1. Radar beam intersecting a gas-solids suspension.

The radar signal is transmitted and then reflected back to the radar antenna by the solids. The relationship between the reflected signal power $P_r$ and $c_v$ is derived from the radar equation. The radar equation [21] expresses the signal power reflected from an individual scatterer:

$$P_r = \frac{P_t G^2 \lambda^2 \sigma_b}{(4\pi)^3 R^4}, \quad (1)$$

where $P_t$ is the peak transmit power, $G$ is the antenna gain, $\lambda$ is the wavelength of the signal, and $\sigma_b$ is the back-scattering cross-section.

The back-scattering cross-section for an individual solids particle with radius $r$ is calculated as:

$$\sigma_b(r) = Q_b(r)\pi r^2 \quad (2)$$

where $Q_b$ is the back-scattering efficiency calculated using the Mie theory formalism for an approximately spherical particle [22]. However, when the radar beam intersects a solids cloud all the individual solids particles within the radar beam contribute to the total back-scattering cross-section, $\sigma_{b,tot}$. Furthermore, multiple scattering effects, i.e., the impact of second-order scattering of photons that do not leave the radar's field-of-view after having scattered once with solids in the radar beam, gain relevance as the solids concentration increases, thereby altering the back-scattered signal. Multiple scattering effects are negligible for sub-wavelength-sized solids when the mean-free path (inverse of the extinction coefficient) between scattering events is larger than the radar's field-of-view (radar footprint) [23]. Theoretically, for approximately wavelength-sized solids, Mishchenko et al. [24] have estimated a solids volume fraction of $2.4 \times 10^{-3}$ as a rough threshold value for validity. The suspensions diagnosed in this work fulfil most often this criterion, as discussed in the *Results* section. Thus, neglecting multiple scattering effects, the total back-scattering cross-section can be expressed as:

$$\sigma_{b,tot} = n_{solids} V_{radar} \int_{r=r_{min}}^{r=r_{max}} f_{PSD}(r)\, \sigma_b(r) dr, \quad (3)$$

where $n_{solids}$ is the solids number concentration (the number of solids particles per unit volume), and $f_{PSD}$ is the PSD function.

As the radar beam is travelling through the gas-solids suspension, it is reflected. It travels back to the radar antenna, experiencing two-way attenuation due to extinction (scattering and absorption). The intensity of the radar beam, $I$, is reduced according to [22, 25]:

$$I(R) = I_0 exp^{-2\int_0^l k_{ext}(l)dl}, \qquad (4)$$

where $k_{ext}$ is the attenuation coefficient and $dl$ is the range interval that the beam has travelled. The attenuation coefficient $k_{ext}$ is related to the total extinction cross-section, $\sigma_{e,tot}$, as $k_{ext} = \sigma_{e,tot}/V_{radar}$. The extinction cross-section for an individual solid particle with radius $r$ can be calculated as:

$$\sigma_e(r) = Q_e(r)\pi r^2, \qquad (5)$$

with $Q_e$ being the extension efficiency of an individual solid particle calculated using the Mie formalism, and, neglecting multiple scattering effects, the $\sigma_{e,tot}$ for the gas-solids suspension is:

$$\sigma_{e,tot} = n_{solids} V_{radar} \int_{r=r_{min}}^{r=r_{max}} f_{PSD}(r)\, \sigma_e(r)dr. \qquad (6)$$

However, when making a compromise between the control of the conducted measurements and the size of the measurement set-up, it should be noted that under experimental conditions, as in this work, the radar beam with radius $r_{radar}$ is not always fully immersed in a solids stream with radius $r_{stream}$ throughout the whole measured distance (i.e., $r_{radar} > r_{stream}$ in some regions, see details in Section 3). This violates the requirement for the applicability of Eq. (4). To correct for this, the first right-hand side term of Eq. (1) is introduced. In addition, a scaling factor $f_{ext}$ is introduced in the attenuation term to scale $k_{ext}$. Furthermore, in a real radar system, several additional factors influence the measurement, such as the radar hardware components (the gain in receiver amplifiers, loss in wires, filters, and antennas) and signal gain and losses introduced by the digital signal processing needed to extract the velocity and range data. As these factors are generally not known with sufficient precision, the radar instrument needs to be calibrated to estimate the absolute values of the solids concentration (for details of the calibration, see Section 3.2.2). In summary, calibration allows the incorporation of all of the above into a single factor, $K$.

With all the considerations mentioned above, Eq. (1) is re-written as:

$$P_r = max\left(\frac{r_{stream}}{r_{radar}}, 1\right) \frac{K\, n_{solids} V_{radar}}{R^4} \int_{r=r_{min}}^{r=r_{max}} f_{PSD}(r)\, \sigma_b(r)dr \; exp^{-2\int_0^l (f_{ext} n_{solids} \int_{r=r_{min}}^{r=r_{max}} f_{PSD}(r)\, \sigma_e(r)dr)dl} \qquad (7)$$

As the last step, the measured solids volume fraction can be calculated as:

$$c_v = n_{solids} V_{solids} \qquad (8)$$

where the solids mean volume is given by:

$$V_{solids} = \frac{4}{3}\pi \int_{r=r_{min}}^{r=r_{max}} f_{PSD}(r)\, r^3 dr. \qquad (9)$$

## 3. Methodology

The following sub-sections describe the procedure for validating the radar measurements of solids concentrations, the experimental set-up for establishing a controlled solids stream, the radar system and its calibration.

### 3.1 Validation procedure

The radar measures the reflected power along a solids stream with mass flow rate $\dot{m}$. For a stable and steady flow of solids, the relationship between $c_v$ and $\dot{m}$ is given by closure of the mass balance:

$$c_v = \dot{m}/(\rho_{part} A_{stream} v_d), \quad (10)$$

where $\rho_{part}$ is the particle density of the solids, $A_{stream} = \pi(d_{stream}/2)^2$ is the cross-sectional area of the solids stream, and $v_d$ is the solids mean velocity.

The level of agreement between the experimental values of $c_v$ derived from radar measurements using Eq. (7) and the values derived from mass flow rate measurements using Eq. (10) is analysed to validate the concentration measurements effected by the radar. The relative error between these values is defined as:

$$E_{rel} = |1 - \frac{c_v \, from \, radar}{c_v \, from \, mass \, balance}| \quad (11)$$

To calculate $c_v$ from the radar signal and from the mass balance closure, the parameters on the right-hand sides of Eq. (7) and Eq. (10), respectively, must be determined. For the mass balance closure, measurements of the mass flow rate, $\dot{m}$, and the solids stream cross-sectional dimension, $d_{stream}$, see Section 3.2.1; and for the radar images used to calculate the solids mean velocity $v_d$, see Section 4. Regarding the radar measurements, determinations of $\sigma_b$ and $\sigma_e$ using Eq. (2) and Eq. (5), respectively, are enabled by the solids specifications presented below (e.g., the PSDs of the solid shown in Figure 4), while determinations of the radar set-up parameters $K$ and $V_{radar}$ are described in Section 3.2.2.

### *3.2 Experimental set-up*

### *3.2.1 Gas-solids flow system*

The experimental set-up was designed and built to establish a controlled, constant, and homogeneously dispersed stream of falling solids, while simultaneously providing the possibility to position the radar beam vertically along such a solids stream. Figure 2a depicts the detailed set-up configuration, indicating the use for radar measurements. The set-up design is inspired by an equivalent set-up reported in the literature [26]. The solids are stored in a first container (storage hopper) that has gaps for the solids to flow down into a second container (discharge hopper), which has a conical bottom with equally distributed holes of 2.8 mm in diameter. The shape and size of the discharge hopper make it suitable to reduce the formation of solids bridges over the holes, which would otherwise cause blockage. The number of open holes in the discharge hopper is used to control the mass flow rate of the solids, while the storage hopper simply guarantees a constant height of the particulate material above the holes of the discharge hopper. In this way, solids discharged from the holes are not subjected to any additional pressure exerted by the weight of piled-up material. After leaving the discharge hopper, the solids pass a metallic funnel in which the solids strings formed through the holes of the discharge hopper are transformed into a single stream of homogeneously distributed solids. The lower opening of the metallic funnel defines the solids stream cross-section. Since the flow of solids is known to entrain air from the surroundings, small gaps are built between the individual hoppers and the funnel to allow air intake from the surroundings and to avoid a pulsing flow of solids stream that would occur otherwise. The stream of falling solids lands onto a 4-mm-thick, high-density polyethylene (HDPE) plate that is tilted at an angle of 45°. The inclination ensures that the solids slide continuously into a solids receiver container, to avoid solids accumulating on the HDPE plate. Since the solids do not flow with the same velocity into the particle receiver as they fall out of the storage hopper (i.e., there are still some solids piling up in the lower part of the HDPE plate), the set-up configuration in Figure 2b was used to measure the solids mass flow rate $\dot{m}$ for each number of holes in the hopper. The distance between the lower funnel opening and the HDPE plate is about 80 cm. The radar is placed ca. 40 cm below and 40 cm to the side of the set-up, aligned towards a flat metal reflector that re-directs the radar beam upwards towards the funnel opening. This implies that the radar beam crosses the HDPE plate and intersects the trajectory of the free-falling stream of solids. Note that the range of the radar is opposite to the falling height of the particles. The range presented in the *Results* section is indicated in Figure 2a. The angle of the HDPE plate with respect to the radar beam minimises reflections from the HDPE plate surface back into the radar. To align the radar beam, the reflected radar signal was maximised along with the range of the solids stream. Measurements were performed for four different particulate materials (as specified below) and five different mass flow rates. To measure the cross-sectional size of the solids stream, $d_{stream}$, photographs of the solids stream were taken with a 10-cm vertical spacing (to avoid distortions at large angles in the photographs' field-of-view). A photograph of a plastic sphere (diameter, 3.2 cm) positioned in the centre of the solids stream was used as a reference scale. In the configuration used for solids mass flow measurements (Figure 2b), the stream of falling solids lands into a container placed upon a precision weighing scale. The mass of collected solids is acquired over time for the different study cases (i.e., different particulate materials and a different number of holes in the discharge hopper). The solids mass flow rate, $\dot{m}$, is calculated from the time derivation of the cumulative mass curve obtained and is shown to remain constant with time.

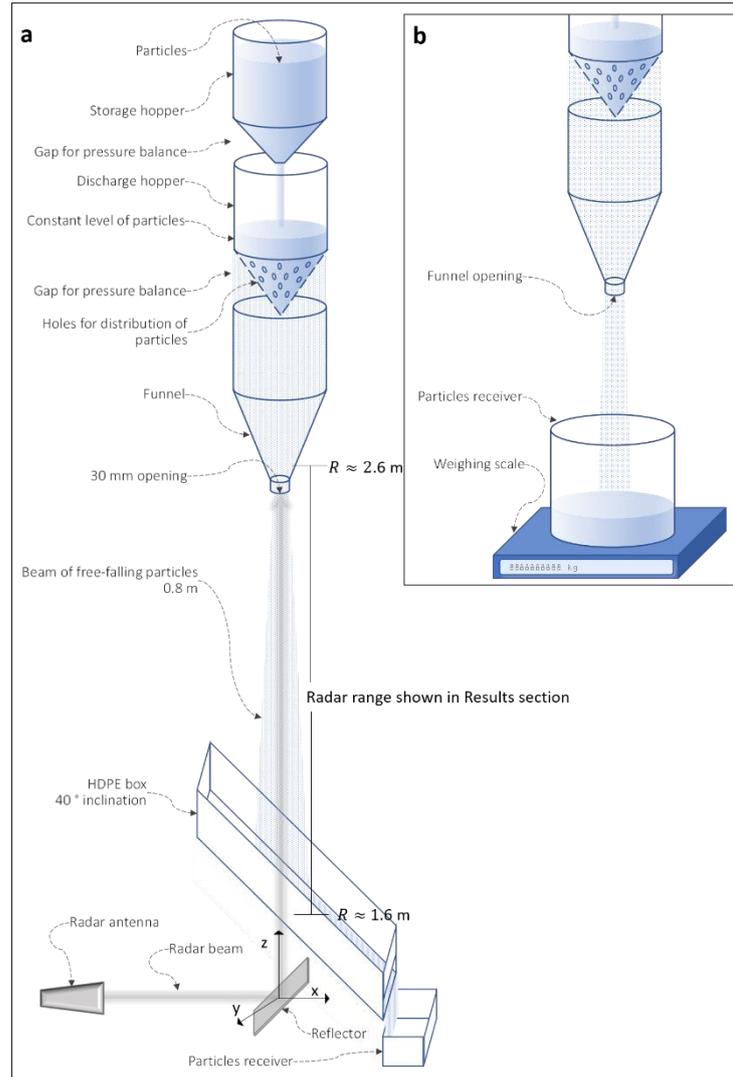

**Figure 2.** *Measurement set-up for: (a) the radar measurement; and (b) determination of the mass flow rate. HDPE, high-density polyethylene.*

Regarding the solids particles, different materials (glass, bronze and sand) are tested to assess the performance levels of the radar measurement for other physical properties of the solids, such as size, shape, and dielectric constant. These materials are regularly used in industrial-scale and laboratory-scale units hosting gas-solids flows. The properties of these materials are summarised in Table 1. The values of the dielectric constant were taken from the literature [27, 28], whereby bronze has an effectively infinite dielectric constant but the value of 1,000 is used for calculations of $Q_b$ and $Q_e$. The particle density was measured by helium porosimetry, while the PSD was measured by light diffraction. Figure 3 shows the cumulative fractions of the PSDs and the corresponding probability density functions. The shape of the different solids is seen in the microphotographs presented in Figure 4.

*Table 1. Summary of the properties of the materials.*

| Material | Density [kg/m$^3$] | Relative dielectric constant at 300 GHz | Shape |
|---|---|---|---|
| Glass (~70%–75% SiO$_2$) | 2,486 | 6 | Spherical |
| Sand (>98% SiO$_2$) | 2,655 | 6 | Irregular |
| Bronze (~10%–20% Sn, 80%–90% Cu) | 8,492 | Infinity (set at 1,000 for the calculation) | Spherical |

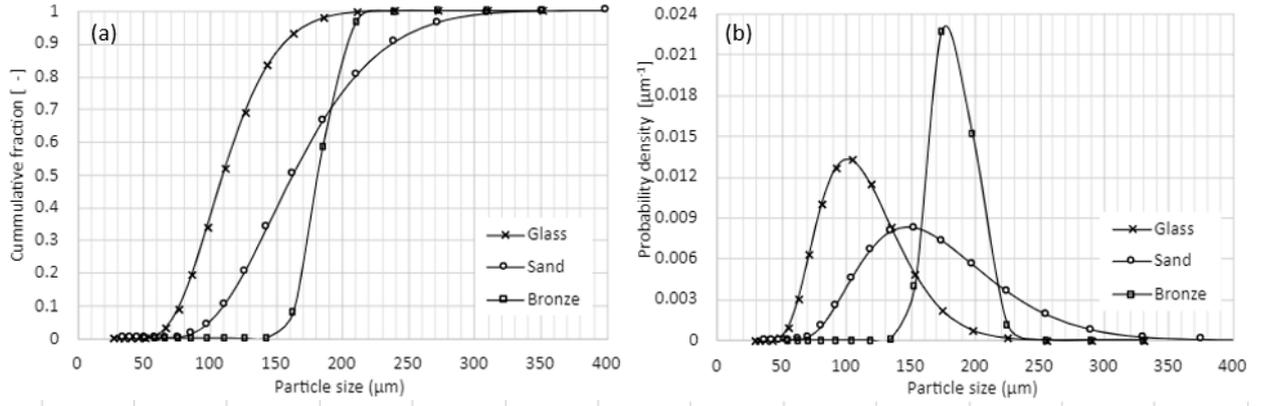
Figure 3. Particle size distributions of the four different types of solids used in this study. (a) The measured cumulative volumetric fraction. (b) The corresponding probability density functions.

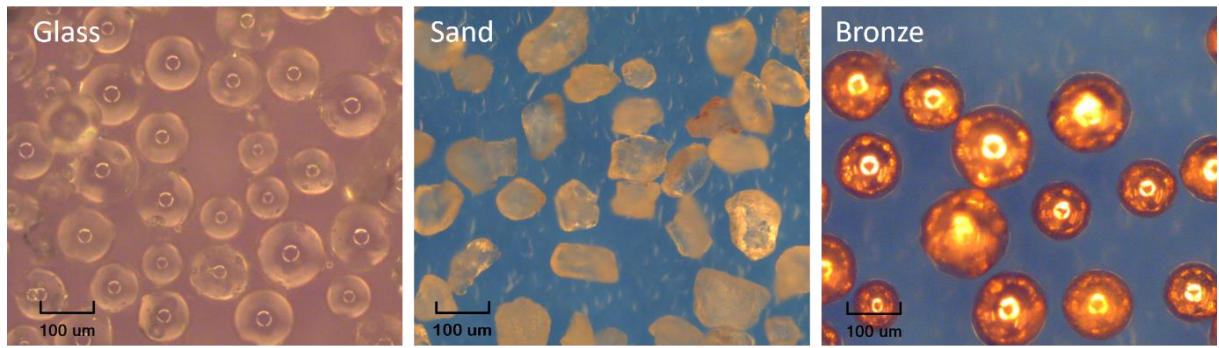
Figure 4. Micrographs of the different types of solids used in the present work. The shapes of the glass and bronze particles are approximately spherical, whereas the sand corns exhibit irregular shapes.

### 3.2.2 The radar system and its calibration

A detailed description of the 340-GHz FMCW range-Doppler radar utilised in this work can be found elsewhere [Tomas2021]. This radar transmits sequences of linearly frequency modulated pulses, and the reflected pulses are processed to find the range (i.e., position relative to the emitter) and velocity of the target solids. For the radar system to become a suitable tool in the characterisation of the dynamics of solids suspensions, the following radar settings must be adjusted:
- the cross-range resolution, $\Delta d$;
- the range resolution, $\Delta R$;
- the range interval, $R_{min}$ to $R_{max}$;
- the velocity resolution, $\Delta v$;
- the velocity interval, $-v_{max}/2$ to $v_{max}/2$, where negative and positive velocities mean that targets are moving towards and away from the radar, respectively;
- the frame rate (setting the measurement speed), $f_{frames}$; and
- the sensitivity towards particle detection (signal-to-noise ratio).

The above-mentioned specifications can be optimised by choosing radar settings and hardware accordingly, where the most important parameters are:
- the centre frequency, $f_c$;
- the pulse bandwidth, $BW$;
- the pulse duration, $t_p$;
- the pulse repetition interval, $PRI$;
- the number of pulses, $n_p$; and
- the receiver bandwidth.

Ultimately, whether or not a particle can be detected by the radar at a certain range depends on the signal-to-noise ratio. The choice of $f_c$ influences how strong the radar signal is reflected. For a constant particle size, the reflected signal power increases with a larger value of $f_c$, i.e., the sensitivity with which smaller particle sizes can be measured increases if all other parameters are kept constant. Furthermore, the resolution of the monitored sample volume is limited by $f_c$ and $BW$, since the cross-range resolution and the range resolution are defined by $\Delta d \approx c_0 R/f_c D$ and $\Delta R \approx c_0/(2BW)$, respectively, where $c_0$ is the speed of light, and $D$ is the diameter of the radar antenna or lens.

The maximum unambiguous radar range is set by $R_{max} = (c_0(PRI - t_p))/2$, with the condition that $t_p$ must be smaller than $PRI$. As reflections from targets that are further away than $R_{max}$ arrive after the next pulse has already been emitted, the radar interprets the target as being closer than it is in actuality. The velocity resolution and the maximum unambiguous velocity are limited by $\Delta v = c_0/(2f_c n_p PRI)$ and $v_{max} = c_0/(2f_c PRI)$. By comparing the expressions for $R_{max}$, $\Delta v$, and $v_{max}$, one can see that the choice of PRI implies a trade-off between maximising $R_{max}$ or $\Delta v$ and $v_{max}$. However, for the rather short ranges of a few metres that are relevant in this work, there is no need for such a trade-off.

Another trade-off that must be considered in this work is between $f_{frames}$ and $R_{max}$, $\Delta v$ and $v_{max}$. The frame rate, $f_{frames}$, is theoretically set by the product $PRI \cdot n_p$. However, before the signal processing can take place, the pulses need to be converted from an analogue into a digital signal by sampling the signals with sampling rate, $f_s$. Larger values of $f_s$ increase the number of samples ($n_s$) needed per converted radar pulse, thereby producing a vast amount of data and, consequently, increasing the signal processing time. However, at the same time, $f_s$ is limiting the maximum range as $R_{max} = f_s c_0 t_p/(2BW)$. Therefore, the chosen value of $f_s$ should be as low as possible, so as to minimise $f_{frames}$ while still resolving the relevant ranges. Furthermore, the receiver bandwidth determines which ranges can be observed by the radar, i.e., the distance at which the observation window is located.

The settings used for the radar system in this work are summarised in Table 2. Note that while the maximum possible sampling frequency of the system is $f_s = 250$ MHz, to reduce the volume of stored data it is reduced by a factor of 4–8 at different stages of the range-Doppler signal processing. Therefore, $\Delta R$ is reduced by a factor of 8 and $R_{max}$ by a factor of 32. The execution time for the pulse train and consecutive signal processing ($f_{frames}$) is set at 0.08 s.

*Table 2. Summary of radar parameters*

| Centre frequency | $f_c$ | 340 GHz |
| --- | --- | --- |
| Bandwidth | $BW$ | 19.2 GHz |
| Range resolution | $\Delta R$ | 8 mm |
| Pulse duration | $t_p$ | 25 µs |
| Number of pulses | $n_p$ | 128 |
| Pulse repetition interval | $PRI$ | 48 µs |
| Number of samples | $n_s$ | 974 |
| Sampling rate | $f_s$ | 250 MHz |

The output data from the radar is a frequency power spectrum, where the measured frequency translates to the range, as described above. The measured power spectrum shows the relative power of the velocity and range and can be transformed to absolute power values with the help of the radar calibration described below. To estimate the parameters $K$ and $V_{radar}$ in Eq. (7), the radar is calibrated using a target with a known radar cross-section and the radar beam pattern is measured at different distances from the radar, respectively. This is achieved by measuring the reflected signal of a metal sphere with diameter $d_{sphere} = 1.27$ cm attached to a two-axis stepper stage. The reference co-ordinate system for the calibration measurement is indicated in Figure 1, replacing the solids-gas suspension with a single metal sphere target. In the first case, the metal sphere is stepped along the propagation direction of the radar beam with ~5-cm increment step. The parameter $K$ is calculated at each position using the equivalence $K = P_r R^4/\sigma_{sphere}$, where $\sigma_{sphere} = \pi (d_{sphere}/2)^2$ is the radar cross-section of the metal sphere when $d_{sphere} \gg 1/(c_0 f_c)$ [29]. Figure 5a shows the calibration measurement, where each circle represents the reflected radar signal of the metal sphere at that position.

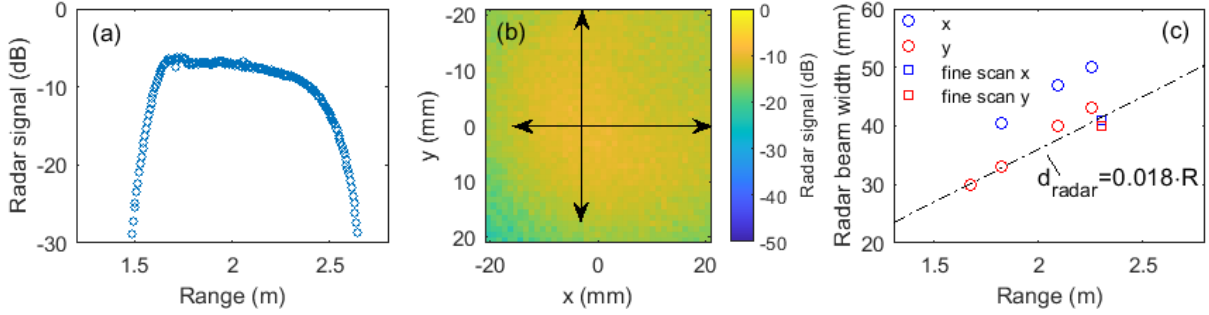

Figure 5. (a) The radar signal reflected by a metal sphere with diameter $d_{sphere}$ = 1.27 cm along the direction of propagation of the radar beam. Each circle corresponds to one measurement point. (b) Radar beam pattern measurement at a distance of 2.3 m from the radar antenna. The scanning area is 41 mm × 41 mm with 1-mm increment. The arrow-headed lines indicate the radar beam-widths ($d_{radar}$) along the *x*-axis and the *y*-axis in relation to the reference system shown in Figure 1. (c) Radar beam-width versus range. The dashed line represents $d_{radar} = R\theta$, with a radar angular beam-width $\theta$ = 0.018 rad.

In the second case, the metal sphere is moved step-wise through the radar beam in the *x* and *y* directions, that is, in a plane perpendicular to the direction of propagation of the radar beam. The radar beam pattern is measured at distances of 1.7 m, 1.8 m, 2.1 m, and 2.3 m from the radar antenna across a scanned area of 64 × 64 mm, with increment step of 4 mm. To enhance the accuracy of the beam pattern measurement, an additional measurement is conducted at 2.3 m with a scanned area of 41 × 41 mm using a finer increment step of 1 mm. Figure 5b shows the radar signal distribution in the *x-y* plane for the fine scan (1 mm increment step) of the radar beam pattern measured 2.3 m from the antenna. The arrow-headed lines connect the half-power points along the *x*-axis and *y*-axis, yielding the radar beam-width, $d_{radar}$. Figure 5c shows $d_{radar}$ versus range. The rough scan (4 mm increment step) indicates that $d_{radar}$ increases with R. The asymmetry between the *x* and *y* directions is due to the larger increment step, which does not allow to discovery of $d_{radar}$ with greater accuracy than±8 mm. The fine scan shows that there is less asymmetry and allows for the calculation of the radar angular beam-width $\theta \approx d_{radar}/R$=0.018 rad, which is used for the calculation of $V_{radar}$.

Lastly, using the same set-up as was used for the calibration, the attenuation of the radar signal by the HDPE plate (see Figure 2) is found to be1 dB. This value was obtained by measuring the signal reflected from the metal sphere, with and without the HDPE plate being placed between the metal sphere and the radar. To take attenuation into account, $K$ is adjusted by a factor of $10^{-0.1}$. Note that the dimension of $P_r$ in Eq. (2) is [W], which is transformed to [dB] according to: Ps [dB] = $10\log_{10}$(Pr [W]).

## 4. Results and discussion

In the first part of this section, the results for the measurements obtained with the lowest mass flow rate of glass beads are presented in detail, to exemplify for the reader the intermediate data obtained. The second section focuses on illustrating the performance of the radar measurements for various mass flow rates of the same solids. Finally, the third section evaluates the radar measurement for the three different tested materials at the lowest values of $\dot{m}$. This simplifies Eq. (7) because the attenuation, and thus the last term, is negligible, such that any possible contribution of multiple scattering effects is avoided.

### 4.1. Glass beads at one given mass flow rate

Figure 6a shows the image (a composite of photographs taken at 10-cm height intervals) of a falling stream of 150–180-μm-diameter glass beads. The mass flow rate for the case illustrated is chosen as 6.5×$10^{-3}$ kg/s, to enhance the contrast between the solids stream and the background in the image. The image shows a homogeneously distributed and stable stream of falling solids exiting the metallic funnel with increasing stream width downstream, which turns into a flickering stream when it is approximately 40 cm away from the funnel. In the flickering section, two parts can be identified: a denser core; and a less-defined dispersed solids cloud in the anulus. These observations agree well with observations reported in the literature [30]. The solids stream width $d_{stream}$ determined from such photographs is exemplified by the data in Figure 6b, where the uncertainty associated with determining the outer edge of the stream by means of digital image analysis is indicated by the gap between the white and blue arrows (indicating the limits of the outer and inner stream regions, respectively). Figure 6b shows also the radar beam cross-sectional size, $r_{radar}$, at different locations. Note that since the radar beam is pointed upwards towards the funnel opening, the radar beam-width increases with distance to the radar antenna, whereas the solids stream width increases in the opposite direction (with falling height from the funnel). This demonstrates that for a certain section

($R > 2$ m as seen in Figure 6b) the radar beam-width is larger than the particle stream width, which is considered in the analysis through the first right-hand term in Eq. (7).

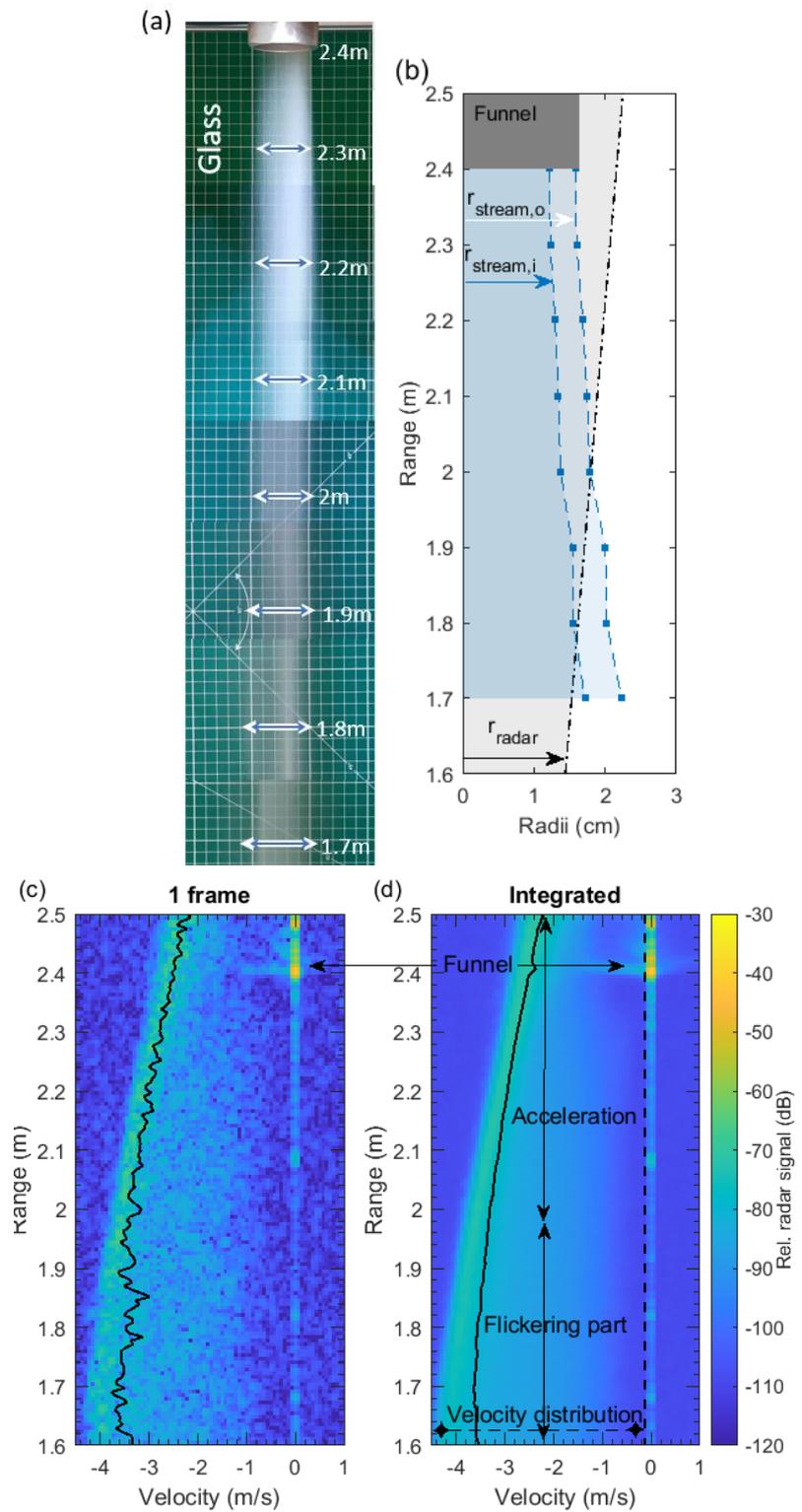

Figure 6. Analyses of a stream of glass beads with mass flow rate of 6.5×10⁻³ kg/s. (a) Photograph of the solids stream. The white and blue arrow-headed lines represent the outer and inner widths of the solids stream and are positioned at 10-cm height intervals. (b) The outer and inner radii, $r_{outer}$ and $r_{inner}$, of the solids stream extracted from (a), together with the radar beam radius, $r_{radar}$ (dashed line), versus range. A radar-measured range-velocity image for a single frame (c) and time-averaged over 15 s (d). The images show the measured radar signal relative to the radar calibration constant. The line indicates the solids mean velocity.

Figure 6, c and d show the corresponding range-velocity images for a single frame and time-averaged data over 15 s, respectively. The data displayed are for the relative radar signal, meaning that the measured signal has been divided by the calibration constant $K$; thus, the relative radar signal strengths can be directly translated to relative solid concentration levels when attenuation is negligible. The images demonstrate that the radar can resolve the range of the solids stream and the corresponding velocity distributions for the different ranges. The blue line in Figure 6d indicates the mean velocity of the solids versus range. Note that for $R>2.45$ m, the radar can even see through the solids stream into the funnel, visualising the solids velocity inside the funnel. At the funnel opening, a distinct local minimum of the mean velocity is observed due to the strong reflection from the opening, which spans a broader interval of velocities (see a detailed explanation in [20]).

Figure 7a shows the profile of the relative radar signal integrated across the measured velocities for each vertical position shown in Figure 6d. For integration, the velocity interval [-0.14, 0] m/s has been excluded, so as to minimise the effect of strong static reflections of the set-up itself, i.e., the funnel opening at $R≈2.45$ m, and some parasitic reflections that can be seen on the zero-velocity line. Since the strong reflections from the set-up itself spill over a broader interval of velocities, the reflections of the funnel remain visible as peaks in the graph. Other than that, the relative radar signal is not affected by any parasitic reflections. Figure 7b shows the upper and lower limits of the solids concentration derived from the mass balance with the profiles derived from the radar measurements, which were calculated based on the relative radar signal in Figure 7a.

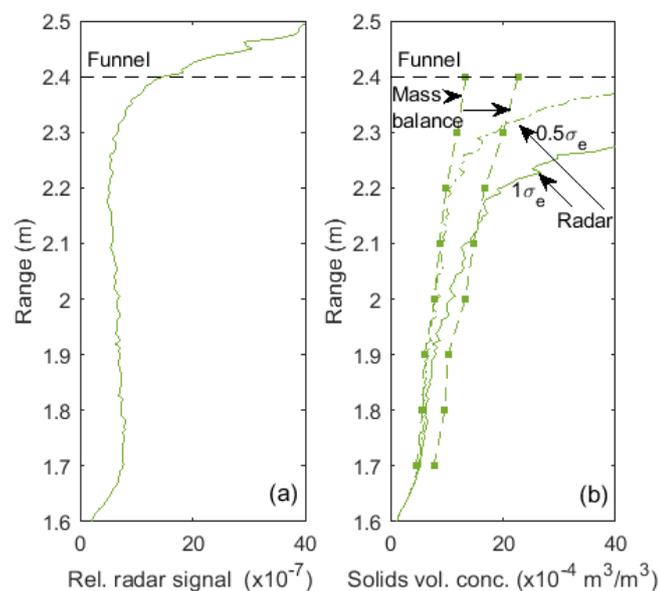

Figure 7. Analyses of a stream of glass beads with mass flow rate of 6.5×10⁻³ kg/s. (a) Velocity-integrated relative radar signal versus range. (b) Solids volume concentration profiles derived from the radar measurement in (a), without (solid line) and with (dashed-dotted line) scaled $\sigma_e$, and from closure of the mass balance (squares-dashed lines).

The solids concentration profile obtained from the radar measurements lies between the upper and lower limits set by the mass balance closure up to $R ≈ 2.1$ m. For $R > 2.1$ m, the range interval within which $r_{radar} > r_{outer}$ (see Figure 6b), the radar measurements yield higher values than the values found via mass balance closure, with increasing disagreement with range being noted towards the funnel. By scaling the extinction cross-section by the factor $f_{ext}= 0.5$ in Eq. (7), the agreement between the radar measurements [dashed-dotted line in Figure 7(b)] and the mass flow measurements is enhanced, with the exception of locations very close to the funnel opening, which could be explained by the parasitic reflection of the funnel.

### 4.2 Glass beads at varying mass flow rates

Investigation of the photographs (see Figure 6a) for the different mass flow rates indicates that the finding of a first homogeneously distributed and stable solids stream that increases in width and thereafter turns into a flickering section approximately 40 cm downstream, holds true for all the mass flow rates tested in this work. Thus, the width of the solids stream and its variation with height, as shown in Figure 6a, is used for all the mass flow rates. Figure 8 shows the range-velocity image for three different mass flow rates of the glass beads. It is clear that higher flow rates result in increased radar signal strengths (which indicate higher solids concentrations). Figure 8 also shows that the solids velocities at a given height increase in magnitude and have broader distribution at higher flow rates. This demonstrates the ability of the radar technique to capture the changes in solids flow dynamics.

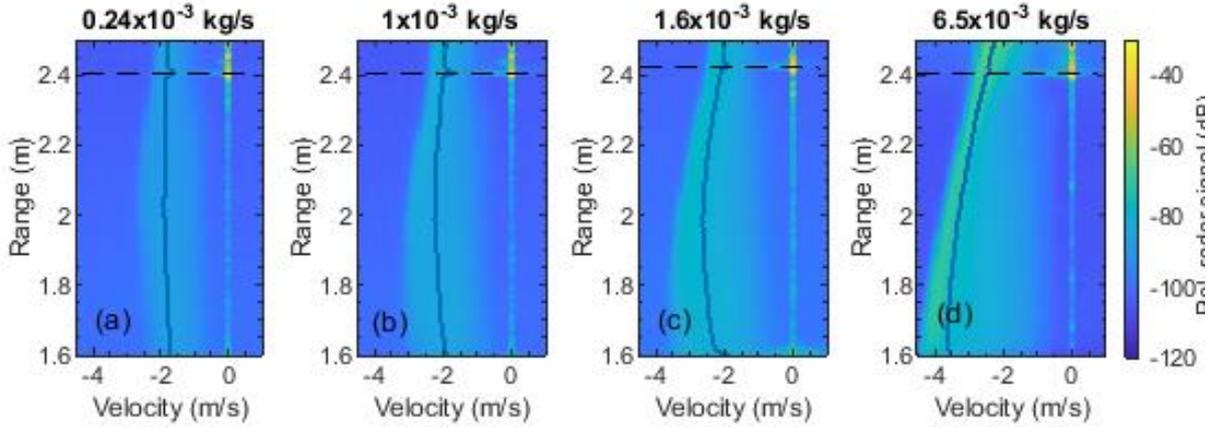

Figure 8. Range-velocity radar images of glass beads with mass flow rates of: (a) 0.24×10⁻³ kg/s; (b) 1×10⁻³ kg/s; (c) 1.6×10⁻³ kg/s; and (d) 6.5×10⁻³ kg/s. The images show the measured radar signals relative to the radar calibration constant. The solid lines indicate the solids mean velocity.

Figure 9a shows the velocity-integrated relative radar signal for mass flow rates ranging from 0.24×10⁻³ kg/s up to 6.5×10⁻³ kg/s and Figure 9, b–d show the comparison between the corresponding radar-measured values of the solids concentration and those derived via the mass balance.

The radar-measured solids concentrations lie within the lower and upper limit for the mass flow rates of 0.24×10⁻³ kg/s and 1×10⁻³ kg/s. For the mass flow rate of 1.6×10⁻³ kg/s, scaling of $\sigma_{ext}$ with the factor $f_{ext} = 0.5$ is required for good agreement with the upper limit. The scaled results (dashed-solid lines in Figure 9, b and c) are also shown for the lower mass flow rates. While the scaling influences the obtained solids concentration profiles for high flow rates (and especially for locations $R$ >2 m), it does not have a significant impact the lowest flow rates because attenuation is negligible. The levels of $c_v$ and the ranges at which the radar system fails to detect solids reflect an interplay between back-scattering efficiency and attenuation. The limit was not reached for the levels of $c_v$ tested here. Theoretically, using the same radar settings presented in this work, assuming a 1-m-long container that is homogeneously filled with glass solids, the reflected radar power should be sufficient to measure at least up to $c_v$ =0.07 m³/m³, noting that the radar settings can be further optimised for the specific application in question.

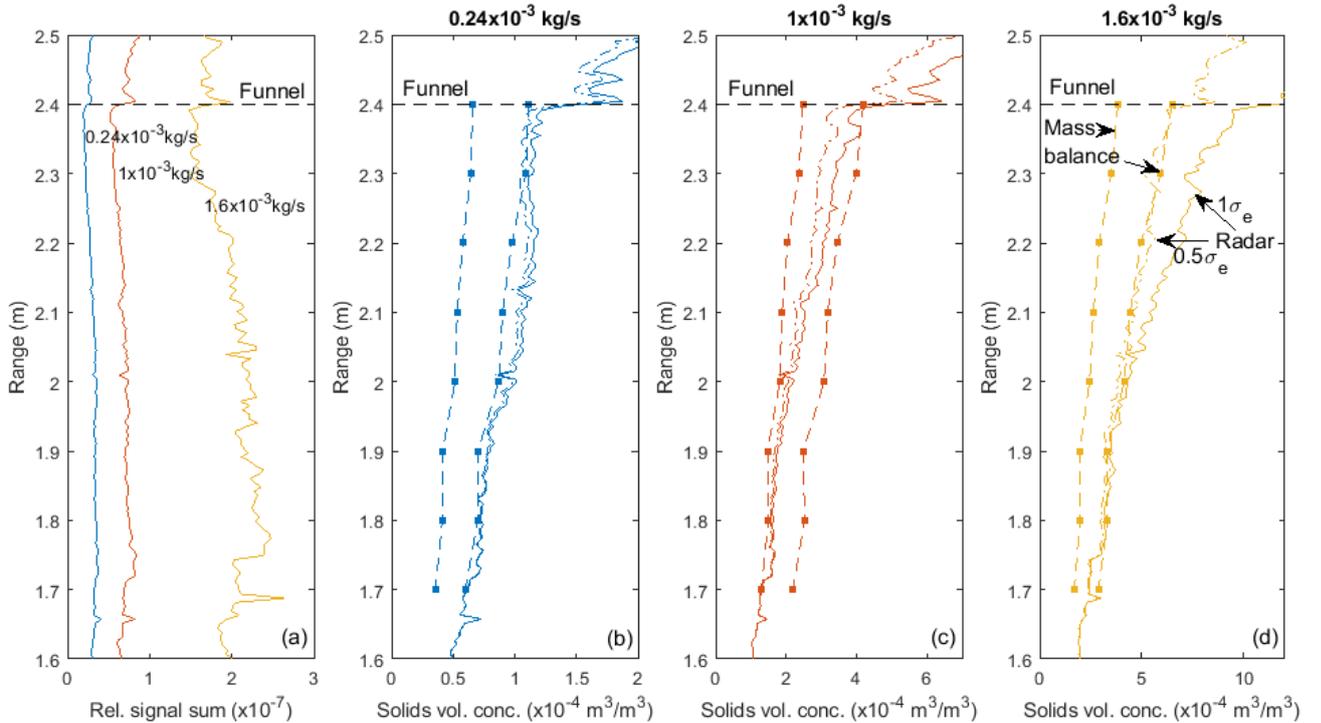

Figure 9. Tests with the glass beads at various mass flow rates. (a) Velocity-integrated relative radar signal versus range. (b–d) Solids concentration profiles defined using the radar measurements in (a), without (solid lines) and with (dashed dotted lines) scaled $\sigma_e$, and from closure of the mass balance (squares-dashed lines).

*4.3 Different particulate materials at low mass flow rates*

Figure 10 shows the range-velocity radar images for the lowest mass flow rates of the glass, sand, and bronze solids. Note that the flow rate for the bronze solids is considerably higher due to its much larger particle density and flowability (owing to greater sphericity) (see Table 1). The plot data for the glass beads and sand do not show any significant differences in terms of signal strength or velocity. Furthermore, the radar signal strength for the bronze spheres is not significantly different, although it is clear that the mean velocity of the bronze particles is higher than those of the sand and glass beads. These results demonstrate the ability of the radar to measure solids flows for a range of materials.

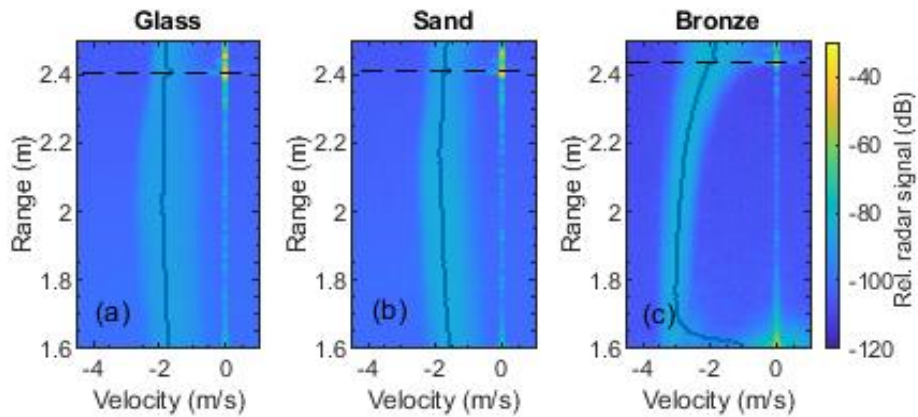

**Figure 10.** Range-velocity radar images for (a) glass beads, (b) sand corns, and (c) bronze beads at mass flow rates 0.24×10$^{-3}$ kg/s, 0.24×10$^{-3}$ kg/s, and 1.9×10$^{-3}$ kg/s, respectively. The images show the measured radar signals relative to the radar calibration constant. The solid lines indicate the solids mean velocity.

Figure 11, a and b show photographs of the solids streams for the sand and bronze spheres, with the arrow-headed lines indicating the extracted solids stream widths. The solids stream widths of the glass beads (see also Figure 6a) and sand are approximately the same. The bronze sphere solids stream has a distinctively different shape, exhibiting much faster diverging stream width with falling height.

Figure 12a shows the velocity-integrated relative radar signals for all the materials tested in this work. Even though sand has the same dielectric constant as glass, sand has a stronger radar signal than glass, which can be explained by its broader PSD with a larger mean particle size. In contrast, the radar signal for bronze is much stronger close to the funnel opening, which is attributed to its large dielectric constant, and it falls off rapidly with decreasing length, which is explained by the rapid growth of the solids stream width and the higher mean velocity, which results in a much-sparser solids beam. At the HDPE plate, the radar signal for the bronze spheres shows a strong peak due to vigorous bouncing of the particles and the fact that few bronze particles penetrate and stick to the plate, thereby adding to the signal.

Figures 9 and 12 show that the solids concentration profiles obtained from the radar measurements show a higher level of agreement with the upper concentration limit derived from the mass balance using $r_{inner}$ as the stream width. This is because the power distribution across the radar beam-width shows its peak power in the centre region (see Figure 3b) and the solids stream has a slightly denser core under these flow conditions (see Figures 6 and 11).

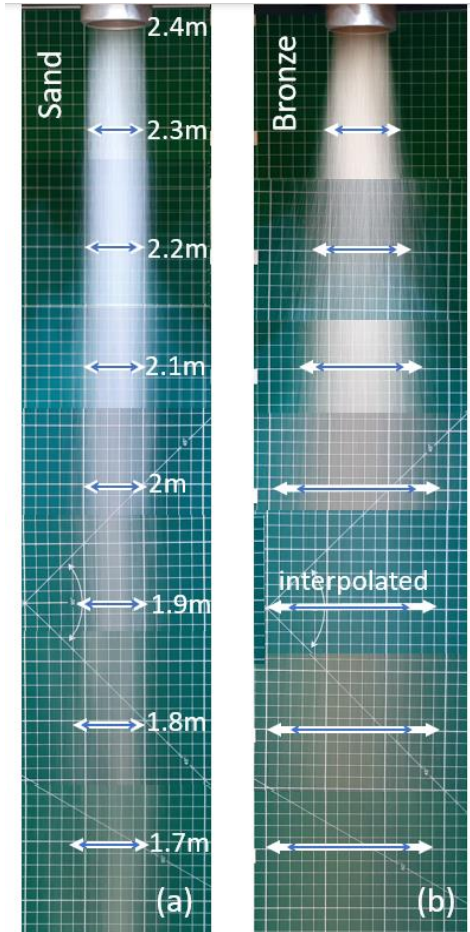

Figure 11. Photographs of the solids stream of: (a) sand; and (b) bronze. Sand is at $\dot{m} = 6.5 \times 10^{-3}$ kg/s, and bronze is at $\dot{m} = 53 \times 10^{-3}$ kg/s.

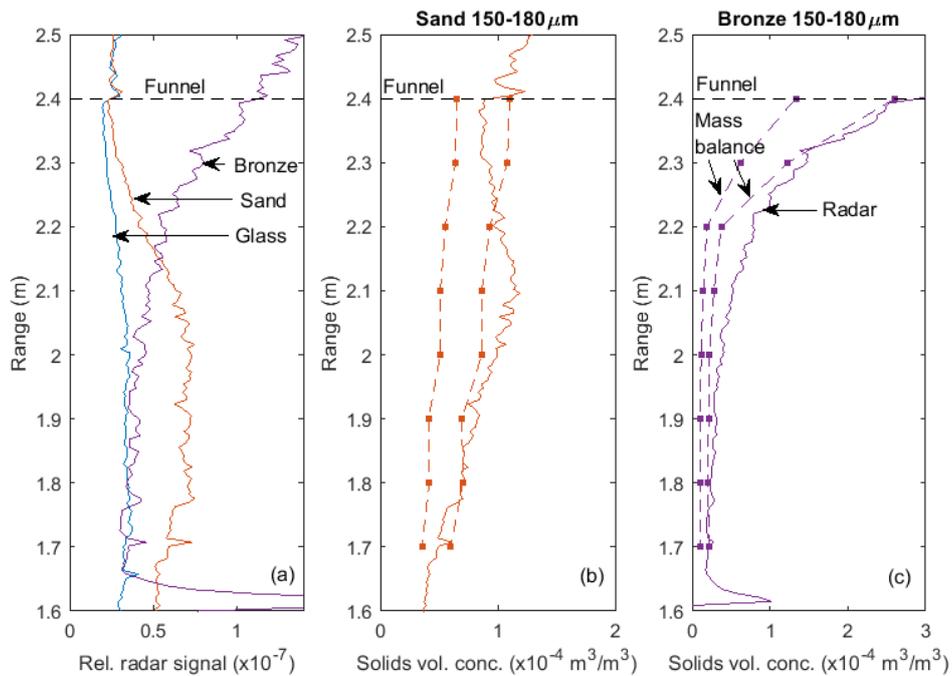

Figure 12. (a) Velocity-integrated relative radar signal versus range for different materials. (b–d) The solids volume concentration profiles defined according to the radar measurements in (a) (solid lines), and from closure of the mass balance (squares–dashed lines).

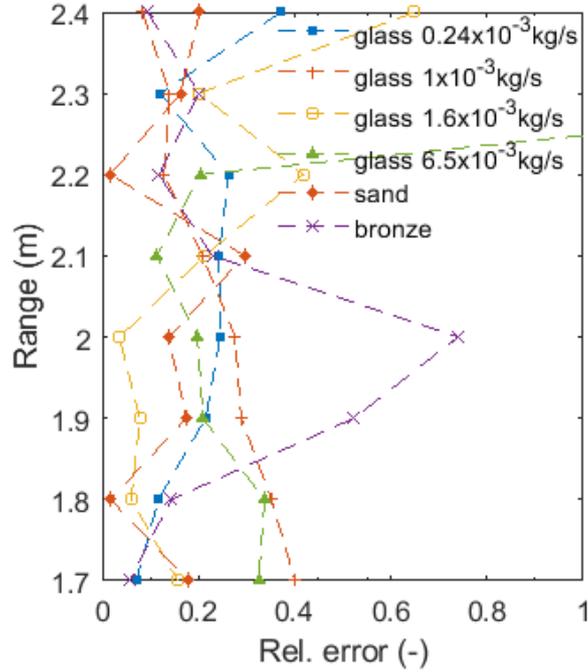

Figure 13. Relative error [Eq. (11)] between the solids concentration profiles from the radar measurements and the upper concentration limit from closure of the mass balance.

Figure 13 shows the relative error between the solids concentration profiles measured by the radar and the upper concentration limit derived via the mass balance for all the mass flow rates and materials. For all the mass flow rates, the extent of agreement is satisfactory, except for two cases in which the relative error is exceptionally large. First, for the highest mass flow rate (6.5×10$^{-3}$ kg/s) at ranges of $R > 2.1$ m, which correspond to the largest solids concentration values tested, the relative error increases dramatically. This is in accordance with the over-estimation of the solids concentration by radar seen in Figure 7b, which is explained in terms of the possible onset of multiple scattering effects and the mismatch of the radar beam-width and the solids stream width (as discussed in Section 4.1). Second, for bronze solids in the distance range of 1.9–2.1 metres, Figure 11b shows that for these ranges, the photographs for evaluating $d_{solids}$ do not have sufficient contrast. Instead, $d_{solids}$ was derived through interpolation, leading to a possible misinterpretation of the bronze stream width. This might be the reason for this larger discrepancy in solids concentration values. Excluding these two cases, the overall relative mean error for all the considered cases is approximately 25% or better.

## 5. Conclusion

This work demonstrates the successful measurement of particulate solids concentration using a radar system. For the tested glass, sand, and bronze with particle diameters in the range of 50–300 μm, the technique was validated with the help of a well-controlled, free-falling solids stream from which the solids concentration was derived through closure of the mass balance. For solids volume concentrations in the range of 0.5×10$^{-4}$ m$^3$/m$^3$ < $c_v$ < 10×10$^{-4}$ m$^3$/m$^3$, the solids concentrations measured by radar are in good agreement with the values estimated from the mass balance closure, showing a mean relative error of approximately 25%. For solids concentrations >5×10$^{-4}$ m$^3$/m$^3$, the theoretical extinction coefficient of the received power by the radar must be decreased if one is to retain a good level of agreement. This can be explained by the onset of multiple scattering effects and a too-small overlap between the radar beam and the solids stream as the latter expands with range and becomes broader than the solids stream. For those materials for which scattering properties are challenging to assess theoretically and for large mass flows, the presented validation method combining mass flow and radar measurements can be used as a calibration method prior to measurement campaigns.

Overall, the range-Doppler based on FMCW radar system is a highly suitable tool for the on-line non-intrusive resolved measurement of the solids flow (concentration as presented in this work, and velocity using the well-established Doppler method). The radar system presented here has a high level of technology maturity for first applications as a diagnostic tool in industry-relevant applications, such as fluidised bed reactors.

*Acknowledgements*
Funding: This work was supported by the Swedish Foundation for Strategic Research (Stiftelsen för Strategisk Forskning) [grant number ITM17-0265, 2019] and the Energy Area of Advance at Chalmers University of Technology.

*Nomenclature*

| Symbol | Units | Description |
|---|---|---|
| $A_{stream}$ | m² | Solids stream cross-section |
| $BW$ | Hz | Radar bandwidth |
| $c_0$ | m/s | Speed of light |
| $c_v$ | m³/m³ | Solids volume concentration |
| $dl$ | m | Length increment |
| $d_{radar}$ | m | Diameter of the radar beam |
| $d_{sphere}$ | m | Diameter of the metal sphere used for calibration |
| $d_{stream}$ | m | Diameter of the solids stream |
| $D$ | m | Diameter of the radar antenna or lens |
| $f_c$ | Hz | Centre frequency |
| $f_{ext}$ | - | Scaling factor of $\sigma_{ext}$ |
| $f_{frames}$ | Hz | Frame rate |
| $f_{PSD}$ | - | Particle size distribution probability function |
| $f_s$ | Hz | Sampling frequency of the analogue-to-digital converter |
| $G$ | - | Antenna gain |
| $I$ | Wm$^{-2}$ | Intensity of radar beam |
| $K$ | Wm² | Radar calibration parameter |
| $k_{ext}$ | m$^{-1}$ | Attenuation coefficient |
| $\dot{m}$ | kg/s | Mass flow rate |
| $n_p$ | - | Number of pulses |
| $n_s$ | - | Number of samples taken per pulse |
| $n_{solids}$ | part./m³ | Solids number concentration |
| $P_r$ | W | Reflected signal power |
| $P_t$ | W | Peak transmit power |
| $Q_b$ | - | Mie back-scattering efficiency |
| $Q_e$ | - | Mie extinction efficiency |
| $r$ | m | Radius of a single particle |
| $r_{radar}$ | m | Radius of the radar beam |
| $r_{stream}$ | m | Radius of the solids stream |
| $R$ | m | Radar range |
| $R_{max}$ | m | Maximum unambiguous radar range |
| $v$ | m/s | Solids velocity |
| $v_d$ | m/s | Doppler velocity |
| $v_{max}$ | m/s | Maximum unambiguous velocity direction |
| $t_p$ | s | Pulse duration |
| $V_{radar}$ | m³ | Radar sample volume |
| $V_{solids}$ | m³ | Solids mean volume |

*Greek letters*

| Symbol | Units | Description |
|---|---|---|
| $\Delta d$ | m | Cross-range resolution |
| $\Delta R$ | m | Range resolution |
| $\Delta v$ | m/s | Velocity resolution |
| $\theta$ | rad | Radar angular beam-width |
| $\lambda$ | m | Signal wavelength |
| $\rho_{part}$ | kg/m³ | Particle density |
| $\sigma_b$ | m² | Single-target back-scattering cross-section |
| $\sigma_{b,tot}$ | m² | Total back-scattering cross-section of solids ensemble |
| $\sigma_e$ | m² | Single-target extinction cross-section |
| $\sigma_{sphere}$ | m² | Radar cross-section of the metal sphere used for calibration |

*Abbreviations*

| | |
|---|---|
| FMCW | Frequency-Modulated Continuous-Wave |
| PSD | Particle Size Distribution |
| HDPE | High-density polyethylene |
| PRI | Pulse Repetition Interval |